\documentclass[%
 reprint,
 amsmath,amssymb,
pra,
]{revtex4-2}

\usepackage{graphicx}
\usepackage{dcolumn}
\usepackage{bm}
\usepackage[mathlines]{lineno}

\usepackage{float}
\usepackage{caption}
\usepackage{subcaption}
\usepackage{upgreek}
\usepackage[english]{babel}
\usepackage{amsmath}
\usepackage{mathtools}
\usepackage{xcolor}

\begin{document}

\preprint{APS/123-QED}

\title{Low-Threshold Lasing with Frozen Mode Regime and Stationary Inflection Point in Three Coupled Waveguide Structure}

\author{Kessem~Zamir- Abramovich$^{1}$}
 \email{kessemzamir@mail.tau.ac.il}
\author{Nathaniel~Furman$^{2}$}
\author{Albert~Herrero-Parareda$^{2}$}
\author{Filippo~Capolino$^{2}$}
\author{Jacob~Scheuer$^{1}$}
\email{kobys@tauex.tau.ac.il}
 \affiliation{$^1$School of Electrical Engineering Tel- Aviv University, Tel-Aviv, Israel}
\affiliation{$^2$Department of Electrical Engineering and Computer Science, University of California, Irvine, California}

\date{\today} 



\begin{abstract}
The frozen mode regime is a unique slow-light scenario in periodic structures, where the flat-bands (zero group velocity) are associated with the formation of high-order stationary points (aka exceptional points). The formation of exceptional points is accompanied by enhancement of various optical properties such as gain, Q-factor and absorption, which are key properties for the realization of wide variety of devices such as switches, modulators and lasers. Here we present and study a new integrated optical periodic structure consisting of three waveguides coupled via micro-cavities and directional coupler. We study this design theoretically, demonstrating that a proper choice of parameters yields a third order stationary inflection point (SIP). We also show that the structure can be designed to exhibit two almost-overlapping SIPs at the center of the Brillouin Zone. We study the transmission and reflection of light propagating through realistic devices comprising a finite number of unit-cells and investigate their spectral properties in the vicinity of the stationary points. Finally, we analyze the lasing frequencies and threshold level of finite structures (as a function of the number of unit-cells) and show that it outperforms conventional lasers utilizing regular band edge lasing (such as DFB lasers). 
\end{abstract}

\maketitle


\section{\label{sec:Inrto}Introduction}

Optical structures operating in the slow-light regime exhibit group velocities that are substantially lower than the speed of light in vacuum. These optical structures exhibit numerous interesting properties; hence, they are highly attractive for many applications. In particular, such structures have been found to exhibit enhancement of properties such as gain, absorption and quality factor (Q factor) ~[\onlinecite{enhancement_low_group_velocity_exp_2006,High_Gain_Slow_light_exp,enhancement_absorption_Slow_light,High_Q_photonic_crystals,Giant_resonances_photonic_crystals2010}]. The frozen mode regime is a special case of slow light, which describes a solution of Bloch wave point with zero group velocity point (stopped light); this point is an outcome of the coalescence of Bloch waves (both eigenvalues and eigenvectors) at a single frequency. The points are called stationary points or exceptional points, and they appear in various types. These types differ in their dispersion properties (the relation between the $k$ vector and the angular frequency $\omega$) in the vicinity of these points \cite{alex2011slowwave}.

In particular, there is a distinct difference between stationary points corresponding to the coalescence of an even and an odd number of modes. The latter are called stationary inflection points~(SIPs). The lowest order of this class corresponds to the coalescence of three eigenvalues, and they are characterized by a cubic dispersion relation:
\begin{align} \label{eq:1}
\omega-\omega_{SIP} \propto (k-k_{SIP})^3
\end{align}

Compared to stationary points with a coalescence of an even number of eigenvalues and eigenvectors, SIPs exhibit several unique properties. More specifically, SIPs are not formed at the band-edge, but rather within the Brillouin zone~(BZ). Since SIPs correspond to the coalescence of three Bloch waves propagating in the \emph{same} direction, they do not form a standing wave. This property renders SIPs interesting scientifically, as well as attractive for various applications involving slow and stopped light. For example, coupling light into and out of an optical structure supporting an SIP is more efficient,
as the excitation of counter-propagating modes can be eliminated \cite{Slow_light_PhC_SIP_efficiency}. This is because in the vicinity of the SIP frequency, the $k$ vector preserves its sign. As a result, counter-propagating waves can be suppressed, contrary to what occurs for stationary points of even order, such as DBEs. Moreover, the slow light regime helps match a slowly propagating mode to a fast mode across the interface of the structure \cite{SIP_DBE_multiple_grating2012}. Furthermore, the SIP resonance has been shown to be remarkably robust to structural disorder and perturbation \cite{frozen_mode_finite_periodic}.

There are many ways to generate stationary points \cite{slow_light_structural_engineering,slow_light_EIT,Ultra_slow_group_velocilty_exp_1999}, one of them being the use of optical periodic structures. In particular, previous studies have presented several periodic structures supporting SIPs, such as waveguides with multiple gratings \cite{SIP_DBE_multiple_grating2012,Nathaniel_Bragg_reflector_grating}, coupled resonators optical waveguides, \cite{DBE_SIP_CoupledRes2017}, three-way periodic microstrip coupled waveguides \cite{SIP_Microstrip_2021}, asymmetric serpentine optical waveguides \cite{asymmetric_serpentine2022}, and coupled transmission lines \cite{SIP_CTL2016}. Besides, previous studies have investigated lasers operating at the frozen mode and stopped light regimes. Lasers operating at a regular band edge~(RBE), such as DFB laser, have been investigated thoroughly. A laser operating at a degenerated band edge~(DBE) was investigated in \cite{DBE_laser2018}. There is limited amount of research on lasers utilizing odd-order stationary points. In Ref.~\cite{unidirectional_SIP_laser_nonreciprocal}, the theory of unidirectional lasers operating in the frozen mode regime has been proposed and studied theoretically. The laser was designed to lase in the vicinity of an SIP, where non-reciprocity was introduced (by means of a magnetic layer) to obtain unidirectional lasing. More recently, a new structure, which also exhibits an SIP, was investigated theoretically as a potential laser near an SIP \cite{SIP_laser2022}. This structure, employing an asymmetric serpentine optical waveguide was shown to exhibit a lower lasing threshold level than that of the same structure operating at an RBE instead.
\begin{figure}[ht]
    \centering
    \includegraphics[scale=0.43]{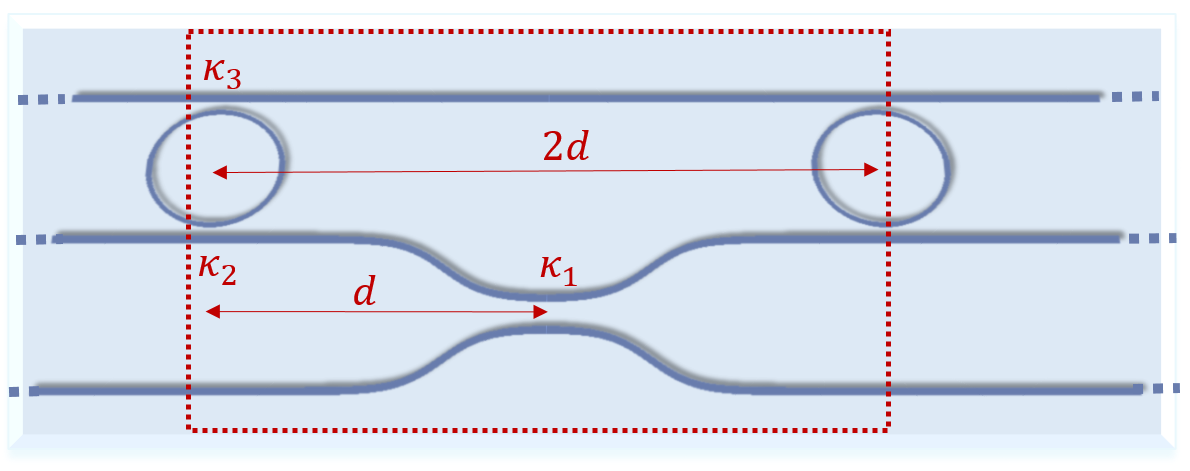}
    \caption{Three periodic waveguides coupled with ring resonator and directional coupler structure. The boundaries of the unit cell are marked by a dashed red line.}
    \label{fig:model}
\end{figure}

In this paper, we propose and study a new, integrated, periodic structure that exhibits SIPs, and can be highly attractive for low-threshold laser applications. The structure consists of a repeating unit cell comprising three parallel waveguides coupled through a directional coupler and a ring resonator, as shown in Fig.~\ref{fig:model}. The central waveguide is coupled to the top waveguide with a ring resonator, and to the bottom one with a directional coupler. By properly setting the parameters of the structure - the resonator radius~($R$), the length of the unit cell~($2d$), and the power coupling coefficients of the coupler and the resonators~($\kappa_1,\;\kappa_2, \;\kappa_3$), it is possible to obtain a dispersion relation exhibiting an SIP, and to control the properties of the structure. We show that this structure is versatile, in the sense that it can support SIPs at various frequencies, depending on the structure parameters. In contrast to many of the previously proposed structures (such as those presented in \cite{asymmetric_serpentine2022}), this structure's unit cell is robust and relatively simple to fabricate. It consists of simple and standard integrated optical elements, whose designed parameters can be readily adjusted and modified to tune the structures characteristics. We also perform a comprehensive analysis of the power transmission/reflection characteristics of this structure in two cases: 1) when all six I/O ports are available and 2) when only one input port and one transmission port are available. This is in contrast to previous studies that only examined the latter case. In addition, we also analyze finite segments of the proposed structure when gain material is incorporated. We calculated the lasing threshold of this geometry, and find that the threshold gain of our structure decreases proportionally to $N^{-3}$, where $N$ is the number of unit cells. We learn that SIP-based devices outperform their RBE counterparts in terms of lasing threshold level. Our results regarding the properties of the threshold gain reinforce the results obtained in previous studies \cite{SIP_laser2022,unidirectional_SIP_laser_nonreciprocal}, especially in the context of SIP-based laser applications. 

The rest of the paper is organized as follows: In Sec.~\ref{sec:dispersion}, we calculate the dispersion relation of the structure by finding a general analytic expression for the transfer matrix of a unit cell. The structure parameters are then optimized in order to obtain SIPs. In Sec.~\ref{sec: Finite_Length}, we numerically calculate the transmission and reflection for a finite structure. We discuss the spectral properties of the structure with three input/output~(I/O) ports, and also with a single input and single transmission port. In Sec.~\ref{sec:gain}, we analyze an active version of this structure, incorporating optical gain material, and in Sec.~\ref{sec:conclusion} we summarize the results and conclude.
The time convention $e^{ i \omega t}$ is used throughout the paper.

\section{\label{sec:dispersion}Dispersion Relation and SIPs}
\begin{figure}[ht]
    \centering
    \begin{subfigure}[b]{0.45\textwidth}
    \centering
        \includegraphics[scale=0.9]{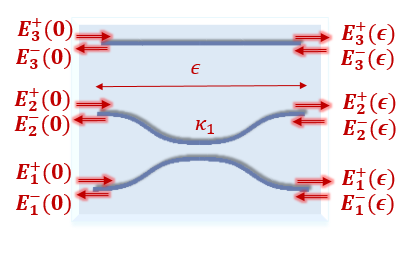}
        \caption{}
        \label{fig:directional_coupler}
    \end{subfigure}
    \begin{subfigure}[b]{0.45\textwidth}
    \centering
            \includegraphics[scale=0.9]{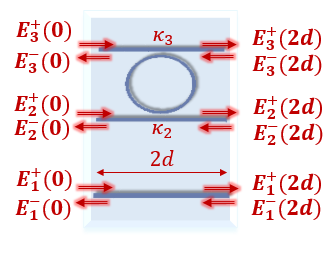}
        \caption{}
        \label{fig:add_drop}  
    \end{subfigure}
     \caption{The two sections of a unit cell: (a) the directional coupler, and (b) the Add- drop multiplexer of length $2d$.}
    \label{fig:two_sections_of_unit_cell}
\end{figure}

We present the derivation of the unit-cell transfer matrix, and obtain the dispersion relation of the eigenmodes in the infinitely-long waveguiding structure. To calculate the Bloch wavenumber dispersion relation of the structure shown in Fig.~\ref{fig:model}, we use the transfer matrix method. For simplicity, we divide the unit cell into two sections: two coupled waveguides with a third separate waveguide seen in Fig.~\ref{fig:directional_coupler}, and an Add-Drop multiplexer of length $2d$ seen in Fig.~\ref{fig:add_drop}. Here we assume the coupling sections are infinitely small (e.g. point coupling, $\epsilon\rightarrow0 $).

Following the unit cell separation, obtaining the transfer matrices for each section is straightforward. The transfer matrix for the three waveguides with a single directional coupler $\kappa_1$  is in Eq.~\ref{eq:matrix_directional_coupler}.
\begin{align} \label{eq:matrix_directional_coupler}
    \begin{bmatrix}
        E_{1}^{+}(\epsilon)\\
        E_{1}^{-}(\epsilon)\\
        E_{2}^{+}(\epsilon)\\
        E_{2}^{-}(\epsilon)\\
        E_{3}^{+}(\epsilon)\\
        E_{3}^{-}(\epsilon)\\
    \end{bmatrix}=
    \begin{bmatrix}
    t_1 & 0 & r_1 & 0 & 0 & 0\\
    0 & t_1 & 0 & -r_1 & 0 & 0\\
    0 & 0 & t_1 & 0 & r_1 & 0\\
    0 & 0 & 0 & t_1 & 0 & -r_1\\
    0 & 0 & 0 & 0 & 1 & 0\\
    0 & 0 & 0 & 0 & 0 & 1\\
    \end{bmatrix} \begin{bmatrix}
        E_{1}^{+}(0)\\
        E_{1}^{-}(0)\\
        E_{2}^{+}(0)\\
        E_{2}^{-}(0)\\
        E_{3}^{+}(0)\\
        E_{3}^{-}(0)\\
    \end{bmatrix}
\end{align}
where $t_1\overset{\Delta}{=}\sqrt{1-\kappa_1}$, $r_1\overset{\Delta}{=}-i\sqrt{\kappa_1}$. Here, $\kappa_1$ is the intensity coupling between the two waveguides. $E_n$ is the electric field phasor in the $n^{\mathrm{th}}$ waveguide.\\\\ 
Similarly, the transfer matrix for the section of the add-drop multiplexer is written in Eq.~\ref{eq:matrix_ADM}, where $T_{1,2},\; D_{1,2}$ are the field transmission functions of the Through and Drop port of two add-drop multiplexers~(ADMs). $T_{1,2},\;D_{1,2}$ are written in Eq.~\ref{eq:Through_Drop_ADM}. Index one ($T_{1},\; D_{1}$) corresponds to an ADM with coupling coefficients $\kappa_3$ at the through port and $\kappa_2$ at the drop port. Index two ($T_{2},\; D_{2}$) corresponds to the opposite case of ADM ($\kappa_2$ at the through port and $\kappa_3$ at the drop port).
\begin{widetext}
{\footnotesize
    \begin{align}\label{eq:matrix_ADM}
        \begin{bmatrix}
        E_{1}^{+}(2d)\\
        E_{1}^{-}(2d)\\
        E_{2}^{+}(2d)\\
        E_{2}^{-}(2d)\\
        E_{3}^{+}(2d)\\
        E_{3}^{-}(2d)\\
    \end{bmatrix}=
     \begin{bmatrix}
    e^{-2i\phi} & 0 & 0 & 0 & 0 & 0\\
    0 & e^{2i\phi} & 0 & 0 & 0 & 0 \\
    0 & 0 & (T_2-\frac{D_1D_2}{T_1})e^{-2i\phi} & 0 & 0 & \frac{D_1}{T_1}\\
    0 & 0 & 0 & \frac{1}{T_2}e^{2i\phi} & -\frac{D_1}{T_2} & 0\\
    0 & 0 & 0 & \frac{D_2}{T_2} &  (T_1-\frac{D_1D_2}{T_2})e^{-2i\phi} & 0\\
    0 & 0 & -\frac{D_2}{T_1} & 0 & 0 & \frac{1}{T_1}e^{-2i\phi}\\
    \end{bmatrix}
    \begin{bmatrix}
        E_{1}^{+}(0)\\
        E_{1}^{-}(0)\\
        E_{2}^{+}(0)\\
        E_{2}^{-}(0)\\
        E_{3}^{+}(0)\\
        E_{3}^{-}(0)\\
    \end{bmatrix}
\end{align}
}
\end{widetext}
\begin{equation}\label{eq:Through_Drop_ADM}
    \begin{aligned}
    T_1 & =\frac{\sqrt{1-\kappa_3}-\sqrt{(1-\kappa_2)(1-\alpha)}e^{-i\varphi}}{1-\sqrt{(1-\kappa_3)(1-\kappa_2)(1-\alpha)}e^{-i\varphi}} \\
    D_1 & =\frac{-\sqrt{\kappa_3\kappa_2}(1-\alpha)^
    {\frac{1}{4}}e^{-i\frac{\varphi}{2}}}{1-\sqrt{(1-\kappa_3)(1-\kappa_2)(1-\alpha)}e^{-i\varphi}} \\
    T_2 & =\frac{\sqrt{1-\kappa_2}-\sqrt{(1-\kappa_3)(1-\alpha)}e^{-i\varphi}}{1-\sqrt{(1-\kappa_2)(1-\kappa_3)(1-\alpha)}e^{-i\varphi}} \\
     D_2 & =D_1  \\
    \end{aligned}
\end{equation} 
$\phi=k_0 n_w \,d$,  $\varphi=k_0 n_r\,  2\pi R$, are phase accumulations in a strait waveguide with length $d$, and a ring resonator with radius $R$, respectively. $n_w$ is the effective refractive index of the straight waveguide, and $n_r$ is the effective refractive index of the curved ring waveguide. $\alpha$ represents intensity loss per revolution inside the resonator ring. For the dispersion relation analysis we assume $\alpha=0$. 
The complete transfer matrix of the unit cell, $\textbf{M}$, is obtained by multiplying the matrices of the two sections. By invoking the Bloch theorem, the dispersion relation of the periodic structure is obtained from Eq.~\ref{eq:Bloch_theorem}. 
\begin{align} \label{eq:Bloch_theorem}
    |\textbf{M}-\textbf{I}  e^{-ik 2d}|=0
\end{align}

where $||$ denotes the determinant operation, and $\textbf{I}$ is the 6x6 identity matrix.

\subsection{\label{subsec: dispersion_separated_SIP}Separated Stationary Inflection Points}

\begin{figure}[ht]
    \centering
    \begin{subfigure}[b]{0.49\textwidth}
    \centering
     \includegraphics[scale=0.585]
    {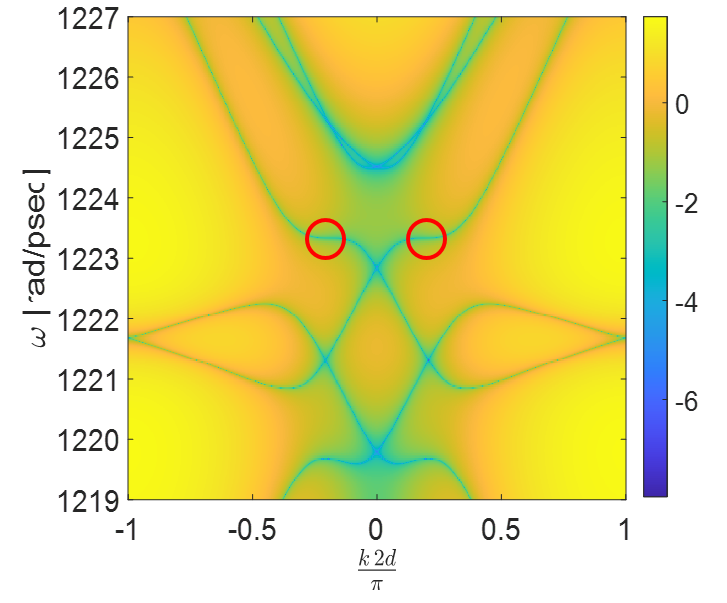}
    \caption{}
    \label{fig:Mat_single_SIP}
    \end{subfigure}
    \hfill
    \begin{subfigure}[b]{0.49\textwidth}
    \centering
     \includegraphics[scale=0.555]    {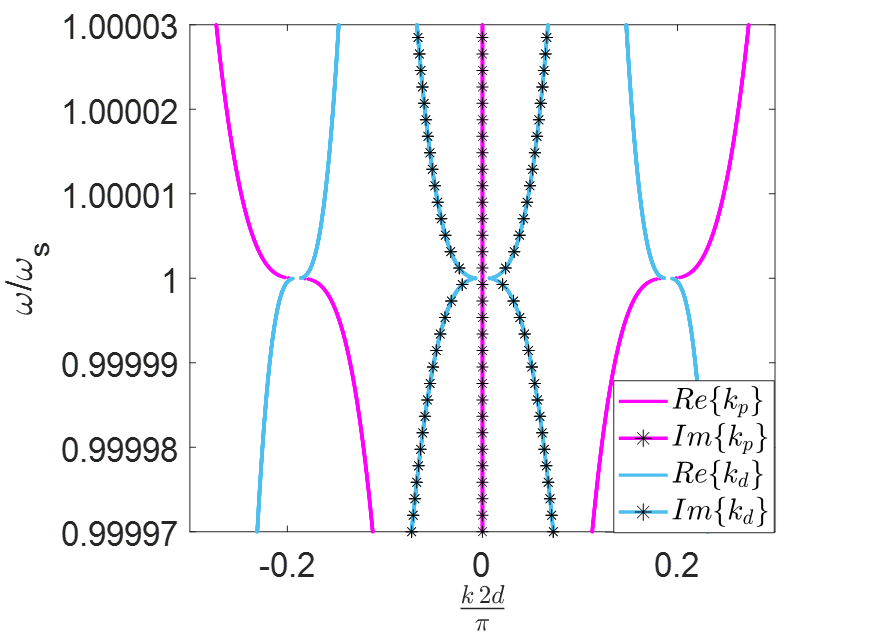}
    \caption{}
    \label{fig:eigenvalues_single_SIP}
    \end{subfigure}
    \caption{`set 1' parameters dispersion relation (a) Logarithmic scale (using base of ten) of the determinant $|\textbf{M}-\textbf{I}\, e^{-ik\,2d}|$, for varying angular frequency and Bloch wavenumber. The SIP point is denoted with a red circle. (b) normalized $k$'s related to the eigenvalues of the transfer matrix $\textbf{M}$ at a range of frequencies near the SIP. The magenta lines indicates the propagating modes, marked as $\{k_p\}$. The light blue lines indicates the decaying modes, marked as $\{k_d\}$. Black-dotted lines and solid lines represent the imaginary and real parts of the normalized wave numbers, respectively.}
    \label{fig:dispersion_single_SIP}
\end{figure}

Eq.~\ref{eq:Bloch_theorem} provides the foundation for analyzing the eigenmodes in the structure depicted in Fig.~\ref{fig:model}, leading to the dispersion relation of an infinitely-long structure for a specific set of parameters ($d$,$R$,$\kappa_1$,$\kappa_2$,$\kappa_3$). Obtaining an SIP requires a specific choice of parameters. Fig.~\ref{fig:Mat_single_SIP} depicts the magnitude of the determinant of the matrix $\textbf{M}-\textbf{I}\, e^{-ik\,2d}$ for the set of parameters: `set 1' listed in Table~\ref{tab:models_parameters}.
\begin{table}
    \caption{Two sets of parameters for studied structure.}
\begin{ruledtabular}
    \begin{tabular}{ccccccccc}
        Parameter & $\kappa_1$ & $\kappa_2$ & $\kappa_3$ & $\alpha$ & $2d [\upmu \textrm{m}]$ & $R [\upmu \textrm{m}]$ & $n_w$ & $n_r$\\
        \hline
        Set 1 & 0.30 & 0.16 & 0.56 & 0 & 44 & 7 & 2.21 & 2.21\\
        Set 2 & 0.20 & 0.42 & 0.73 & 0 & 20 & 4 & 3.42 & 3.42\\         
        
    \end{tabular}
    \label{tab:models_parameters}
\end{ruledtabular}
\end{table}
The blue-green lines in the figure, indicate magnitude close to zero, i.e. solutions of Eq.~\ref{eq:Bloch_theorem}, thus indicating the $k-\omega$ dispersion relation. The red circles denote a potential SIP at $\lambda\approx1.54 \, \upmu \textrm{m}$, related to frequency $\omega_s=\frac{2\pi\,c}{\lambda }$. Note that at this point we obtain $\frac{\partial\omega}{\partial k}=0$, and a third-degree polynomial in its vicinity, which is indicative of a frozen mode regime. Nevertheless, this does not guaranties an SIP, as the coalescence of both eigenvalues and eigenvectors still needed to be verified. 

To verify that this is indeed a stationary point of order three, we consider the eigenvalues of the transfer matrix $\textbf{M}$. Any input field in this structure can be spanned by a basis of six propagating and evanescent waves which are the eigenvectors of $\textbf{M}$. At a given frequency, the eigenvalues of $\textbf{M}$ are designated as $\{\gamma\}_{i=1}^6$, where the relation between the eigenvalues and the corresponding wavenumbers $k$ is given by Eq.~\ref{eq:eigenvalues}. 
\begin{align}\label{eq:eigenvalues}
    \gamma= e^{-i k \, 2d}
\end{align} 
A stationary point of third order (i.e., an SIP) is characterized by the coalescence of three eigenvalues and three eigenvectors. Fig.~\ref{fig:eigenvalues_single_SIP} plots the dispersion relation of the structure (normalized complex wavenumber $\frac{k\, 2d}{\pi}$ vs. angular frequency). At each frequency, there are six Bloch modes with six eigenvalues. The magenta lines in Fig.~\ref{fig:eigenvalues_single_SIP} indicate propagating modes. The solid magenta lines correspond to the real parts of the eigenvalues, while the doted ones indicate the imaginary parts. These eigenvalues are purely real,  and the real parts of the eigenvalues match exactly the green lines near the SIP in Fig.~\ref{fig:Mat_single_SIP}. The light blue lines in Fig.~\ref{fig:eigenvalues_single_SIP} indicate the non-propagating (decaying) modes. Similar to the propagating modes, the solid and dotted lines indicate the real and imaginary parts of the wavenumbers, respectively. Note that there are four decaying modes: two that are complex conjugates. In other words, two modes have eigenvalues with a positive real part, and the other two with a negative real part. From Fig.~\ref{fig:eigenvalues_single_SIP} we identify the SIP frequency as the frequency at which three of the wavenumbers coalesce into a single real value. It corresponds to the frequency marked by a red circle in Fig.~\ref{fig:Mat_single_SIP}, $\omega=\omega_s$. Due to reciprocity, two SIPs are obtained (at the same frequency), corresponding to two triply- degenerate counter-propagating modes. 

We emphasize that an SIP is a condition where both eigenvalues and eigenvectors coalesce. However, Fig.~\ref{fig:eigenvalues_single_SIP} only shows the coalescence of the eigenvalues. The coalescence of the eigenvectors is verified through the coalescence parameter that was defined in \cite{SIP_Microstrip_2021,asymmetric_serpentine2022} for the SIP (see Appendix~A for details). Appendix~A also describes the structure design and optimization process used for obtaining the parameters in Table~\ref{tab:models_parameters}. Consequently, it can be inferred that Fig.~\ref{fig:eigenvalues_single_SIP} verifies that the marked points in Fig.~\ref{fig:Mat_single_SIP} are indeed SIPs, meaning that $\omega_s=\omega_{SIP}$.

\subsection{\label{subsec:Dispesion_almost_overlap}Almost-Overlapping Stationary Inflection Points}

\begin{figure}[ht]
    \centering
    \begin{subfigure}[b]{0.49\textwidth}
    \centering
     \includegraphics[scale=0.58]
    {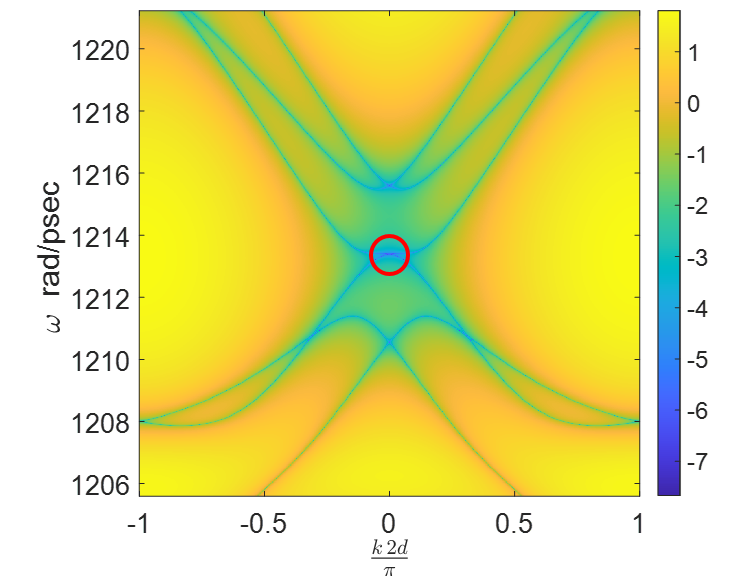}
    \caption{}
    \label{fig:Mat_double_SIP}
    \end{subfigure}
    \hfill
    \begin{subfigure}[b]{0.49\textwidth}
    \centering
     \includegraphics[scale=0.55]
    {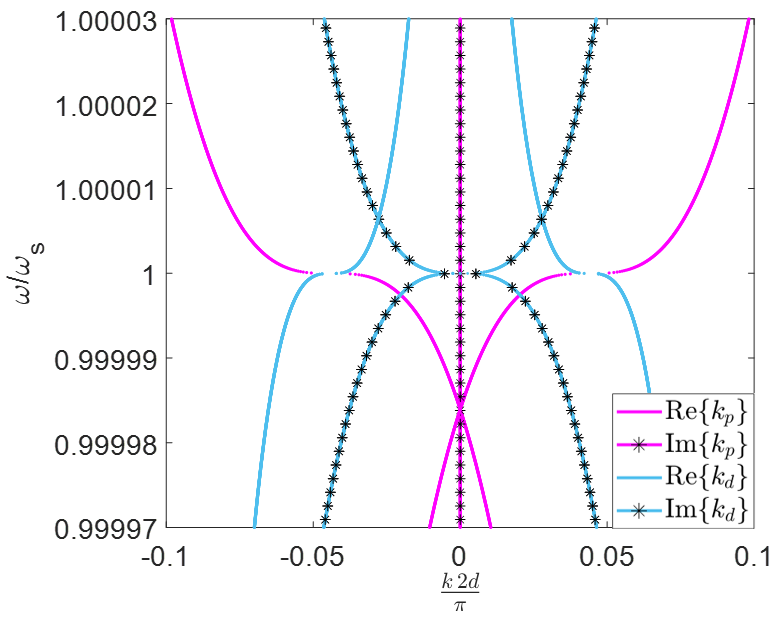}
    \caption{}
    \label{fig:eigenvalues_double_SIP}
    \end{subfigure}
    \caption{`set 2' parameters dispersion relation (a) Logarithmic scale (using base of ten) of the determinant 
    $|\textbf{M}-\textbf{I}\cdot e^{-ik\,2d}|$, , for varying angular frequency and Bloch wavenumber. The SIP is marked with a red circle (b) normalized $k$'s related to the eigenvalues of the transfer matrix $M$ at a range of frequencies near the SIP. The magenta lines indicate propagating modes, marked as $\{k_p\}$. The light blue lines indicate decaying modes, marked as $\{k_d\}$. Black-dotted lines and solid lines correspond to the imaginary and real parts of the normalized wave numbers, respectively.}
    \label{fig:dispersion_double_SIP}
\end{figure}

The proposed structure can exhibit SIP dispersion properties in a range of frequencies and locations in the BZ, controlled by the design parameters.
Modifying the parameters ($d$,$R$,$\kappa_1$,$\kappa_2$,$\kappa_3$), depicted in Fig.~\ref{fig:model}, enables us to generate an SIP in different locations in the dispersion relation. This section considers the set of parameters designated as `set 2' in Table~\ref{tab:models_parameters}. As in Sec.~\ref{subsec: dispersion_separated_SIP}, Fig.~\ref{fig:Mat_double_SIP} plots the colormap of the determinant of $\textbf{M}-\textbf{I}\, e^{-ik\, 2d}$. Fig.~\ref{fig:eigenvalues_double_SIP} depicts the Bloch wavenumbers of the transfer matrix as a function of frequency. At first glance, it seems that a flat band with zero group velocity is formed at the center of the Brillouin Zone~(BZ) with wavelength $\lambda\approx1.55\, \upmu \textrm{m}$, represented with a red circle. We define the frequency related to this wavelength as before, $\omega_s$. In a scenario of a flat band, the transfer matrix of the structure should exhibit a degeneracy of six eigenvalues related to $k=0$. However, the eigenvalues depicted in Fig.~\ref{fig:eigenvalues_double_SIP} indicate that this so-called "flat band" consists, in fact, of two very close SIPs located at both sides of the BZ center. This should not be surprising as it is well known that an intersection of different spectral branches at $k=0$ is only possible for a 1D periodic structure that exhibits glide plane symmetry \cite{Higher_symmetries_theory,Glide_symmetry_microwaves,Glide_symmetry_photonic_crystals,Glide_symmetry_holey_structure}. However, as the structure depicted in Fig.~\ref{fig:model} does not exhibit glide symmetry, the two SIPs cannot fully intersect at the center of the BZ and therefore cannot support degenerate SIPs. 

Thus, we understand that the studied structure can support the formation of SIPs over large domains in the BZ, and at different wavelengths. The frequencies and wavenumbers of such SIPs can be controlled by modifying and optimizing the set of parameters that describe the structure. In Sec.~\ref{subsec: dispersion_separated_SIP} we showed an SIP that is located at $\frac{k\,2d}{\pi}\approx0.2$, which makes this SIP separated from its backward propagating counterpart. In this section we described an SIP that is formed almost at the center of the BZ, at $\frac{k\,2d}{\pi}\approx0.05$. This shows that this structure is incredibly versatile, since it can generate SIPs at various locations in the BZ. 

\section{\label{sec: Finite_Length}Finite Length Structure}
\subsection{\label{subsec:finite_open}Single Input - Multiple Output Transmission and Reflection}

The dispersion relation and frozen mode regimes discussed in Sec.~\ref{sec:dispersion} are obtained only for infinitely-long structures. However, any realistic structure is finite in length, thus leading to resonances with optical properties that may differ a bit from those of the infinitely-long structure. Resonances in such a realistic finite-length structure may not support a ``perfect'' frozen light mode. However, as more unit cells are added, the closer the SIP resonance is to the SIP frequency, and the smaller the group velocity is at the resonance. Thus, the longer the structure, the closer the characteristics of the propagating waves in the structure become to those expected at an SIP. Therefore, we expect to obtain enhancement of properties, such as the Q factor, by using finite-length structures with a sufficient number of unit cells.

In order to investigate the transmission and reflection properties of a finite-length structure, we analyze a waveguide with length $l=2d\, N$ where $N$ is the number of unit cells. Any resonance response of such structure in the vicinity of an SIP is expected to be enhanced due to the small group velocity. In addition, the larger the number of unit cells, the closer the resonance frequency to the SIP. Recall that the structure consists of three parallel waveguides, (see Fig.~\ref{fig:model}) with three I/O ports. Consequently, the structure can be excited in many ways. We analyze the power transmission and reflection properties, for the case where an input field excites only the upper waveguide (port 3) from the left. We use the transfer matrix approach to obtain the intensity of the fields exiting from the three left-hand side waveguides (i.e. reflection), and from the three right-hand side waveguides (transmission) of the structure. Eq.~\ref{eq:T/R} defines the transmission and reflection power, which are denoted as $T_m^2$ and $R_m^2$ respectively. The plus and minus signs indicate forward and backward propagating fields, and the subscript $m$ denotes at which port the field is calculated. The number in the brackets is the point on the propagation axes where the field is calculated. Fig.~\ref{fig:two_sections_of_unit_cell} shows the port numbers. 
\begin{equation}
    \begin{matrix}
        T_m^2=\left|\frac{E_{m}^{+}(l)}{E_{3}^{+}(0)}\right|^2 &
        R_m^2=\left|\frac{E_{m}^{-}(0)}{E_{3}^{+}(0)}\right|^2 &
        m=1,2,3
    \end{matrix}
    \label{eq:T/R}
\end{equation}

\begin{figure}[ht]
    \centering
    \begin{subfigure}[b]{0.49\textwidth}
    \centering
         \includegraphics[scale=0.46]{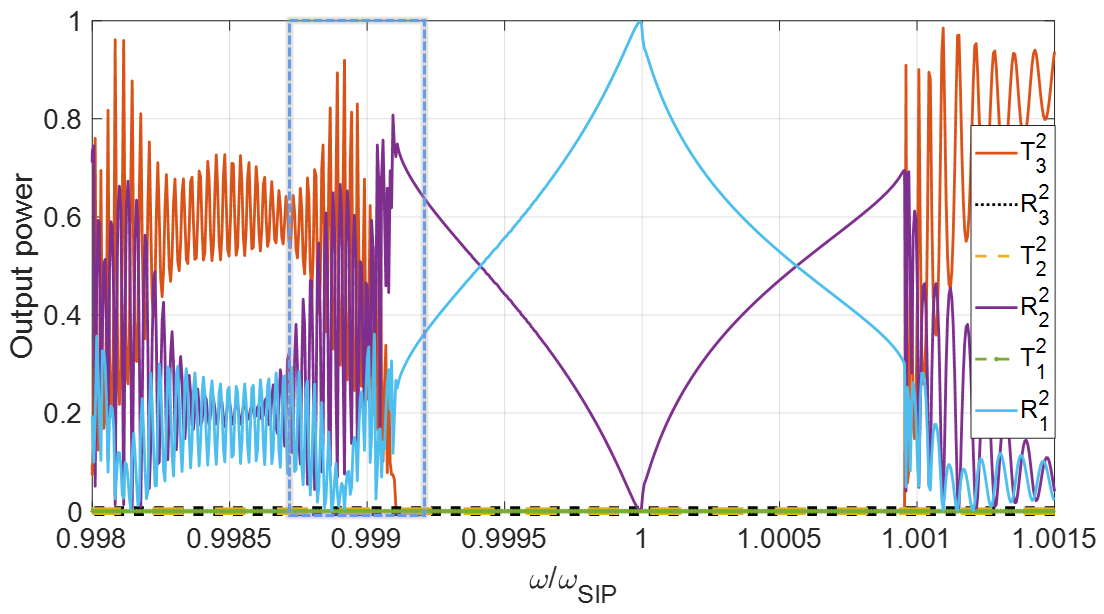}
    \caption{}
    \label{fig:T_R_finite_model_open}   
    \end{subfigure}
    \begin{subfigure}[b]{0.49\textwidth}
    \centering
        \includegraphics[scale=0.55]{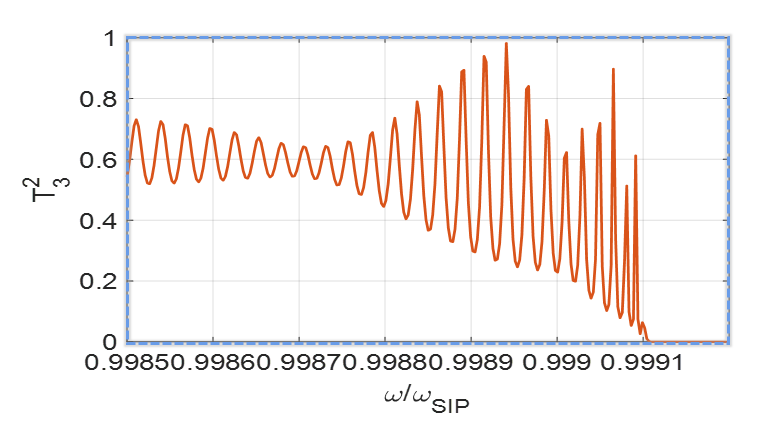}
        \caption{}
        \label{fig:T_port3}
    \end{subfigure}
    \caption{(a) Transmission and reflection from six ports of a finite-length waveguide structure with 60 unit cells. (b) Transmission from input port 3 to output port 3, near the SIP frequency. The parameters in `set 1' are used.}
\end{figure}
Fig.~\ref{fig:T_R_finite_model_open} shows the transmission and reflection spectra of a finite-length structure with $N=60$, excited through the upper waveguide input (port 3). In this figure, the SIP frequency obtained for `set 1' is designated as $\omega_{SIP}$. It can be seen that the transmitted signal through port 3 and those reflected through ports 1 and 2 exchange power as a function of frequency, while the signals at the other ports are zero. At frequencies around the SIP, the spectral profiles in the transmission and reflection ports become more oscillatory and exhibit narrower peaks. This is clearly seen in Fig.~\ref{fig:T_port3}, which is a zoom-in of $T^2_3$ (marked in the black rectangle in Fig.~\ref{fig:T_R_finite_model_open}). At $\omega \approx 0.999\omega_{SIP}$, the fast oscillations in the spectral responses stop and a resonance in the reflection from port $R^2_1$ is formed, accompanied by zero transmission through ports $T^2_3$ and a complementary decrease in $R^2_2$. At $\omega_{SIP}$, all the power is reflected through the lower waveguide ($R^2_1$). The observed increase in the density of the spectral peaks in the vicinity of the SIP indicates that the group velocity is indeed smaller close to the SIP frequency. It also implies that by introducing optical gain into the structure, it should be possible to obtain large optical amplification and even lasing.        

\subsection{\label{subsec:finite_closed}Single Input - Single Output Transmission Port}

\begin{figure*}[ht]
    \centering
    \includegraphics[scale=0.45]
{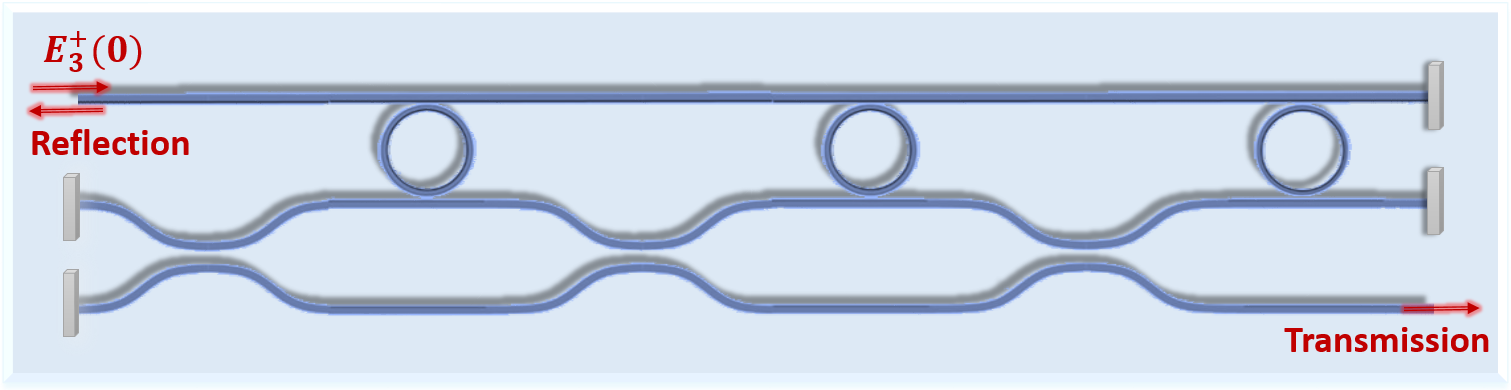}
    \caption{Finite-length model, closed with reflective mirrors. This arrangement has one input port and one transmission port.}
    \label{fig:closed_model}
\end{figure*}

In order to obtain an SIP, it is necessary for the structure to support at least three modes in each direction, thus necessitating the three-path waveguide arrangement. However, for any practical application such as high-gain amplifier or laser, it is advantageous to have a single input and two outputs, reflection (from the input port) and transmission from a second port. This can be achieved by closing the I/O ports at the end of the (finite) structure as shown in Fig.~\ref{fig:closed_model}: Input ports 1 and 2 and output ports 2 and 3 are closed by introducing reflectors which can be implemented as distributed Bragg reflectors~(DBRs). A DBR is a natural choice for implementing an integrated waveguide reflector. In this scenario, we chose the reflectors to exhibit perfect reflectivity. Different DBR designs facilitate high reflectivity as well as control over the reflected spectral band  \cite{DBR_waveguide_fab,MEMS_illustration_DBR_waveguide}. The closed structure possesses two ports which can be considered as I/O for optical amplification purposes or two output ports for lasing (similar to a Fabry-Perot laser). Consequently, the reflection and transmission ports of the structures are at port 3 (left) and at port 1 (right), respectively. The choice between different waveguides for the I/O ports stems from the fact that an SIP formation necessitates coupling between forward and backward propagating waves. The position of the DBRs ensures that the input signal propagates through all the waveguides in both directions. Nevertheless, additional configurations for ``closing'' the structure (i.e. obtaining single input and single output) are possible. The study of such configurations is beyond the scope of this paper. In this finite-length model, we assume no losses, therefore due to energy conservation the sum of the transmission and the reflection powers equals one. 

\begin{figure}[ht]
    \centering
    \includegraphics[scale=0.5]{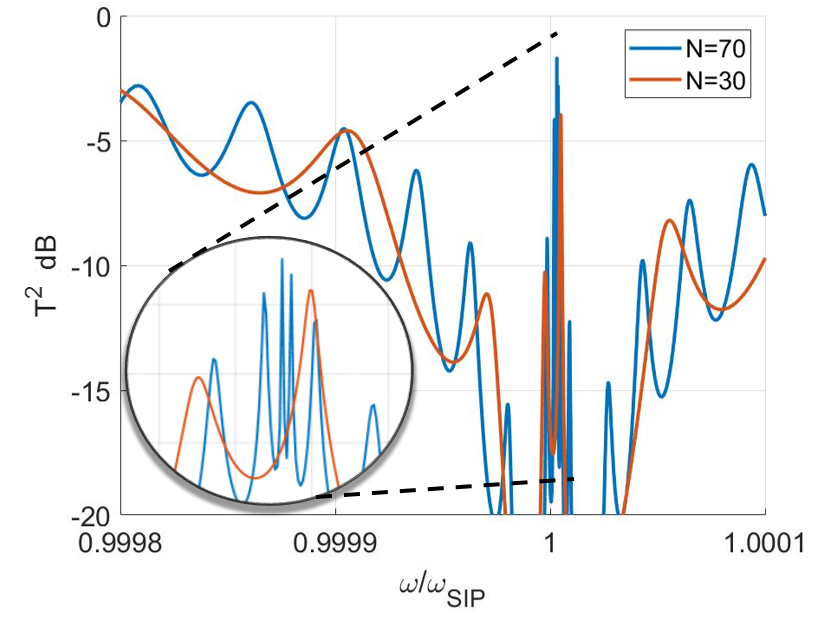}
    \caption{Transmission of closed model with mirrors, for 30 and 70 unit cells.}
    \label{fig:T_R_finite_closed_model}
\end{figure}

Fig.~\ref{fig:T_R_finite_closed_model} plots the transmission spectrum $T^2$, for two structure lengths of 30 and 70 unit cells. There are several things to note in the transmission spectrum of the device. The transmission exhibits oscillations similar to a Fabry-Perot~(FP) cavity (compared to the spectral properties of the `open' structure shown in Fig.~\ref{fig:T_R_finite_model_open}). This result is not surprising, since the closing of some of the ports in the structure introduces feedback that couples between forward and backward propagating waves (as in an FP cavity). However, in the vicinity of the SIP frequency, these oscillations corresponding to the FP resonances become dense and sharper (higher Q factor). This is attributed to the slow-light effect in the vicinity of the SIP which effectively reduces the local free spectral range~(FSR) of the FP cavity. The resonances of the structure consisting of $N=70$ unit cells are denser and sharper than those of the $N=30$ device. This is due to the longer FP cavity, which, similar to the slow-light effect, reduces the FSR. This effect is particularly strong near the SIP frequency (shown in the inset of Fig.~\ref{fig:T_R_finite_closed_model}), where the resonances of the longer structure are much sharper than those of the shorter one due to the combination of the longer device and slow-light effect.   

\section{\label{sec:gain}Analyzing Gain Properties}
The sharp resonances associated with high quality factor and the SIP frequency, indicate that such a structure could be useful for the realization of low-threshold lasers \cite{DBE2016,DBECoupledRes2018,SIP_laser2022}. To identify the lasing threshold of the structure we introduce an imaginary part to the refractive index of the resonators and waveguides, and consider the dependence of the threshold level on the number of unit cells. The lasing condition is a singular point, which is characterized as a pole in the transmission spectral response. In other words, at the lasing threshold, the output power calculated using the transfer matrix, approaches infinity regardless of the input intensity level, as shown in the book \cite{book_semiconductors_lasers}, chapter 5. In practice, gain saturation effects, which are often not considered in a linear transfer matrix analysis, limit the actual output power. Nevertheless, such effects do not affect the lasing threshold.

\begin{figure}[ht]
    \centering
    \begin{subfigure}[b]{0.49\textwidth} 
        \includegraphics[scale=0.49]{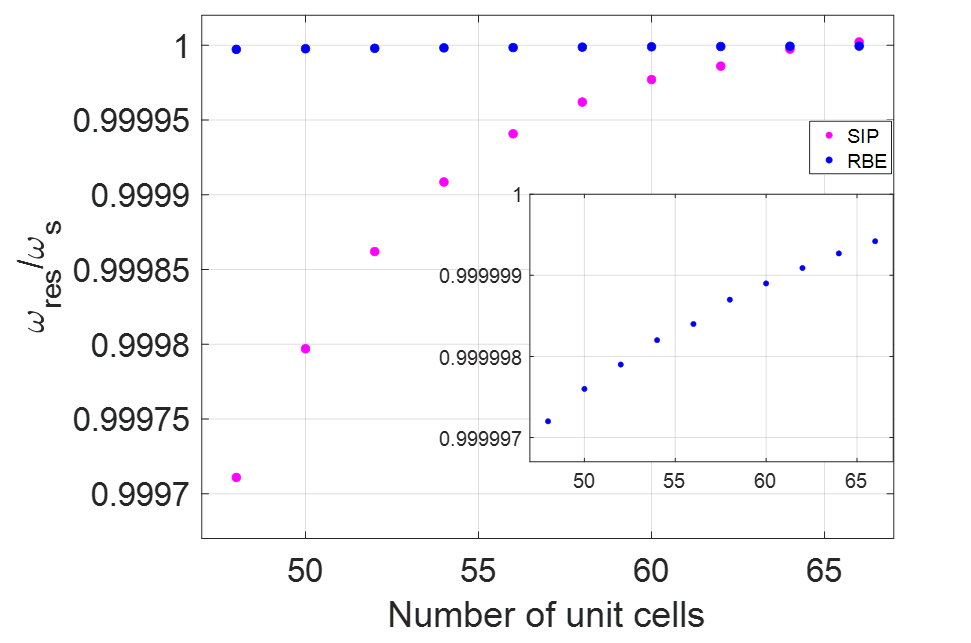}
        
        \caption{}
        \label{fig:threshold_frequency_vs_unit_cells}
    
    \end{subfigure}
    \hfill
    \begin{subfigure}[b]{0.49\textwidth}
           \includegraphics[scale=0.51]{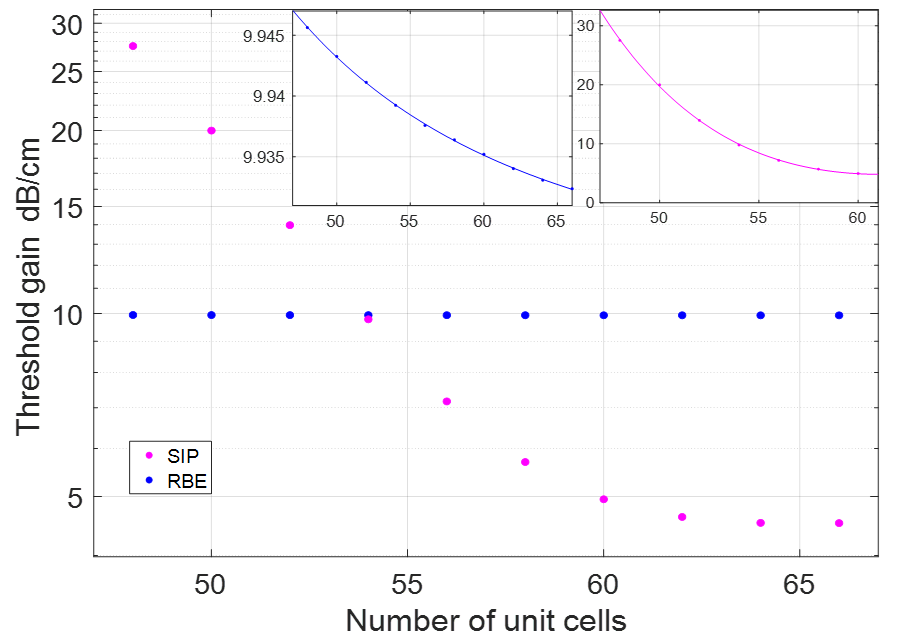}
        \caption{}
        \label{fig:threshold_gain}
    \end{subfigure}
    \caption{Dependence of the threshold gain in the waveguides, and the lasing frequency versus the number of unit cells of the structure. (a) The angular resonance frequency $\omega_{res}$, which is the closest to the stationary point {$\omega_{s}$}, as a function of the number of unit cells. $\omega_s$ refers to the stationary point frequency, either $\omega_{SIP}$ (for the magenta dots)  or $\omega_{RBE}$ (for the blue dots).  Magenta points: change in resonance frequency normalized by the SIP frequency in our structure. Blue points: change in resonance normalized by the RBE frequency in the CROW. (b) The threshold gain for RBE in CROW model, and SIP in the proposed model versus the number of unit cells. The y-axis is in logarithmic scale. Magenta points: SIP threshold gain, Blue points: RBE threshold gain. The two insets show polynomial fitting of the SIP (magenta) and the RBE (blue). Note that the y- axis in those figures is in linear scale. }
\end{figure}

As the structure supports many resonances that can potentially lase, we consider the resonance which is closest to the SIP frequency because it is the one that retains the SIP properties the most. As the group velocity in the vicinity of this frequency is lowest, the intensity buildup at this frequency is expected to be enhanced substantially and exhibit the lowest threshold level. We continue to focus on the SIP obtained with `set 1' parameters. Fig.~ \ref{fig:threshold_frequency_vs_unit_cells} depicts the resonance (i.e., the lasing) angular frequency (denoted by magenta dots) as function of the number of unit cells. The resonance frequency approaches $\omega_s$, which corresponds to $\omega_{SIP}$ frequency for the magenta points, as the number of the unit-cells increases. 

As a comparison, we also calculate the threshold level of a structure exhibiting a regular band edge (RBE) of a CROW structure \cite{CROW2004} comprising ring resonators with radius $R=7 \mu \textrm{m}$ (which is the same as the one in `set 1'  Table~\ref{tab:models_parameters} here considered) coupled to each other with the coupling coefficient $\kappa_r =\kappa_2$ from set 1. The unit cell length of the CROW is $l=4R$, which is smaller than the unit cell length of our structure. Any realistic structure exhibits losses that should be included in the calculation of the lasing threshold. Thus, we introduce a loss parameter of $\alpha=0.01$ to the ring resonators, equivalent to a 1 percent power loss per revolution. 


Although the comparison is not perfect, it does allow us to get a better understanding of the lasing properties of a device operating in the vicinity of an SIP. The process of calculating the threshold gain for a finite structure is detailed in Appendix B. 

Fig.~\ref{fig:threshold_frequency_vs_unit_cells} also shows the lasing frequency of a CROW structure as a function of the number of unit-cells, $N$ (blue dots). The lasing frequencies are normalized to either $\omega_{SIP}$ or to $\omega_{RBE}$, according to the relevant structure. Note that in this case $\omega_{RBE}\approx 1224 (\textrm{rad/ps})$. It can be seen, that the number of unit cells $N$ has more impact on the lasing frequency of the SIP supporting structure than on that of the RBE. However, although the lasing frequency of the RBE supporting structure seems to be independent of $N$, this is not the case. The inset of the figure shows a zoom-in on the lasing frequency of the RBE structure, indicating that it also approaches $\omega_{RBE}$ as $N$ increases. In both cases, the dependence of the resonances on $N$ can be described as $\omega_{th} \propto \alpha+\beta N^{-1}+\gamma N^{-3}$.
Fig.~\ref{fig:threshold_gain} depicts a comparison between the threshold gain (in the waveguides) near the SIP (in pink) of the structure depicted in Fig.~\ref{fig:closed_model}, and that corresponding to the RBE in the CROW. The parameters of 'set 1' in Table~\ref{tab:models_parameters} have been used in the comparison. The threshold gain is presented as a function of the number of unit cells in each structure. The scale of the $y$ axis is in logarithmic scale. There are a couple of important points to be noted: first, the lasing threshold gain of the SIP supporting structure decreases rapidly with the number of unit cells, reaching a minimal level of $\approx4.5$ dB/cm at $N=64$, and second is that the lasing threshold saturates for longer structures. We attribute the threshold gain saturation to the fact that the loss in structure is also proportional to the number of unit cells. Thus, for a sufficiently long structure, this loss mechanism dominates all other mechanisms in the structure (e.g. output coupling loss), and the lasing threshold is obtained when the gain becomes equal to the loss at each unit cell. In contrast, the dependence of the threshold gain on $N$ for the RBE case seems to be approximately constant. However, zooming in on the RBE lasing threshold $N$ dependence (shown as the left inset in Fig.~\ref{fig:threshold_gain}) indicates that this is not the case. The threshold gain near the RBE of the CROW is not constant but rather decreases slowly with $N$, at a slope that is substantially slower than that of the SIP supporting structure.  

Second, by fitting the two calculated curves (see insets of Fig.~\ref{fig:threshold_gain}) we find that the threshold gain dependence on $N$ in both cases can be described by a $3^{rd}$ order polynomial in $N^{-1}$: $a+b \,N^{-1}+c\, N^{-3}$. This dependence agrees with previous results \cite{SIP_laser2022,DBE_SIP_CoupledRes2017,alex2011slowwave}. The main difference between the curves is the value of the fitting coefficients $b$ and $c$ corresponding to the dependence on $N^{-1}$ and $N^{-3}$ respectively. In the CROW structure we find that $|b_{RBE}|\approx 10^{-1} \textrm{dB/cm}$ and $c_{RBE}\approx 2\cdot 10^4 \textrm{dB/cm}$, while in the SIP supporting structure we find that $|b_{SIP}|\approx 2\cdot 10^{-4} \textrm{dB/cm}$ and $c_{SIP}\approx2 \cdot 10^7 \textrm{dB/cm}$. There is a  substantial difference (of orders of magnitude) between these coefficients, which has a dramatic impact on the properties of the two stationary points (the SIP and RBE). The decrease of the lasing threshold as a function of $N$ is significantly faster for the structure supporting an SIP. This is seen by the larger fitting coefficient of the $N^{-3}$ term and the smaller coefficient of the $N^{-1}$ term for that structure. Consequently, the dominant dependence of the lasing threshold of the SIP structure and the CROW is $N^{-3}$ and $N^{-1}$, respectively. It should also be noted that although the threshold value of the SIP structure is larger than that of the CROW for $N<54$, once the structure length exceeds 54 unit cells, the SIP lasing threshold becomes smaller. Although a direct comparison between the structures is difficult because they are very different from each other, the overall trend indicates the advantages of operating near a SIP. Clearly, a laser operating at an SIP can potentially exhibit a lower lasing threshold than that of conventional lasers.   

\section{\label{sec:conclusion}Conclusion}
In this paper we introduced a new periodic structure composed of three parallel waveguides that are coupled to each other by ring resonators and directional couplers. We calculated the dispersion relation of the structure by means of the transfer matrix method, and showed that a proper choice of the structure parameters (coupling coefficients, ring radius and the length of the unit cell), leads to the formation of SIPs in the dispersion relation. We also showed that it is possible to control the frequency and wavenumber of the SIPs. Moreover, we studied the properties of finite-length structures (i.e. with a finite number of unit cells). We calculated the spectral transmission and reflection at each port, and studied their properties in the vicinity of the SIP frequencies. We then studied the transmission and reflection properties of a closed structure, exhibiting only two ports, by introducing reflectors at the other four ports. This is only one of the possibilities of closing such a structure, and other closed-structure configurations might lead to different spectral transmission and reflection. Specifically, we focused on two structures with $N=30$ and $N=70$ unit cells. We found that longer structures yield sharper resonances exhibiting higher Q-factors. We attribute this to the fact that the longer the structure, the closer the resonance gets to the zero group-velocity point in the dispersion relation. Finally, we analyzed the ability of this structure to serve as a laser by introducing optical gain into the structure. Specifically, we focused on the dependence of the lasing threshold gain on the number of unit cells, $N$. We compared the lasing threshold gain at the SIP frequency and that of a structure supporting a second-order stationary point (i.e., RBE). We found that although the dependence of the threshold gain on $N$ is similar in both cases and can be described by a $3^{rd}$ order polynomial in $N^{-1}$, the threshold level of SIP-supporting structure decreases faster with increasing $N$. Thus, we conclude that this structure is highly attractive for low-threshold laser devices.       
\appendix
\setcounter{secnumdepth}{0}
\section{Appendix A: Optimization Process}

This appendix describes the design and optimization process of the SIP-supporting structure. For this design, there are five independent parameters that need to be determined (and optimized): the power coupling coefficients  ($\kappa_1,\kappa_2,\kappa_3$), the ring radius ($R$) and a length parameter, which is related to the length of the unit cell ($d$). There is an additional degree of freedom, which is the position of the directional coupler (in the longitudinal direction) between the middle and the lowest waveguide. Here, we choose to position it at the center of the unit cell, equally distanced from the two micro-rings. Nevertheless, this is just a specific choice. {Changing the position of the directional coupler brakes a mirror symmetry, in the z direction inside the unit cell, which might allow us to get another SIP, because it is a non-symmetric stationary point. Note that this structure does not exhibit any symmetry in the vertical axes, which is a requirement for obtaining an SIP}.

Since there are many parameters needed to be determined, a numerical optimization approach is required. To determine the design parameters we minimize the coalescence parameter in Eq.~\ref{eq:D_h}, introduced in \cite{2019exceptional_Dh}. This parameter quantifies the degree of coalescence between a number of complex vectors, determined by the order of the stationary point -$SP_{order}$. For example, in the case of an SIP, the degree of coalescence of three eigenvectors of the transfer matrix $\{\phi_k\}$, having the same eigenvalue, is minimized. When the angles between all of the eigenvectors approach zero (i.e. all the eigenvectors are parallel), the coalescence parameter approaches zero as well, i.e., $D_h\rightarrow 0$. At the start of the optimization process, the length of the unit cell and the radius of the micro-ring are chosen randomly. The coupling coefficients are then varied manually by means of trial and error until a potential SIP is formed in the dispersion relation at a certain wavelength $\lambda_p$. Then, the coupling coefficients are optimized numerically in an attempt to minimize $D_h$ for the wavelength $\lambda_p$. Here we employ the Nelder-Mead simplex algorithm for the optimization \cite{fminsearch_alg}. Our objective function is: $f(\{\phi_k\})=D_h\left(\{\phi_k\}\right)+10^2\cdot (degeneracy\left(\gamma\right) -degeneracy_{desired})$. The term $D_h$ corresponds to the coalescence parameter of a group of eigenvectors having the same eigenvalue $\gamma$. The \emph{`degeneracy'} parameter corresponds to the degeneracy (i.e. algebraic multiplicity) of the eigenvalue $\gamma$, related to the group of eigenvectors $\{\phi_k\}$. The `$degeneracy_{desired}$' parameter is the desired order of the degeneracy of the eigenvalue $\gamma$ (e.g. $degeneracy_{desired}=3$ for an SIP). 
Therefore, the use of this objective function favors the formation of $degeneracy_{desired}$ identical eigenvectors and eigenvalues.  

\begin{widetext}
\begin{align}
    D_h=\frac{1}{{SP_{order}\choose 2}}\;\;\sum_{\mathclap{\substack{\;\;\;\;\;\;m=1,n=1\\ m> n}}}^{SP_{order}} \sin(\theta_{n,m})\phantom{~~~~~~~~~~~~} \cos(\theta_{n,m})=\frac{\langle \phi_n | \phi_m \rangle}{\| \phi_n \| \| \phi_m \|}
    \label{eq:D_h}
\end{align}
\end{widetext}

\begin{figure}[ht]
    \centering
    \includegraphics[scale=0.6]{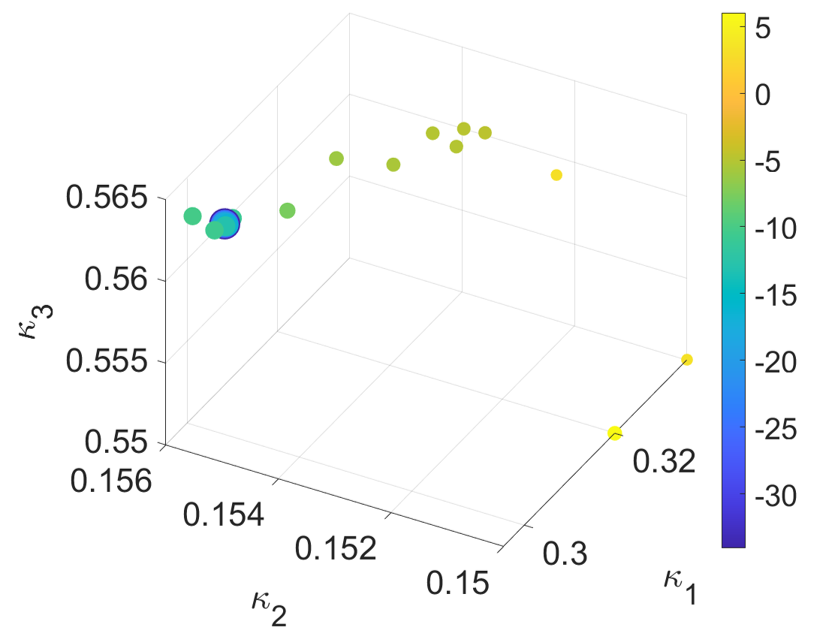}
    \caption{Convergence of optimization process. The $x,y,z$ axes denote the intensity coupling coefficients.  The size of the dots and their color indicate the value of $D_h$ in dB. The bigger and darker the dots are the smaller $D_h$ is.}
    \label{fig:optimiztion_process}
\end{figure}

Fig.~\ref{fig:optimiztion_process} depicts the convergence of the optimization process that yielded the coupling coefficients of `set 1'. In this case, $\lambda_p$ is the wavelength of the SIP $\lambda\approx 1.54\upmu$\textrm{m}, and the ring radius and the unit cell length are defined in Table~\ref{tab:models_parameters}. The dots in the figure indicate the position in the 3D coupling coefficients space $\{\kappa_1,\kappa_2,\kappa_3\}$ of the structure during the optimization process. The size and the color of the dots, indicate the value of the coalescence parameter. The small yellow dot located at $\{\kappa_1,\kappa_2,\kappa_3\}=\{0.32,0.15,0.55\}$ indicates $D_h$ in the initial point (obtained manually as described above) of the optimization process. At this point, $D_h>0.3$, indicating that the eigenvectors have not coalesced yet. The dots show the progress of the optimization process, indicating the trajectory in the coupling coefficients space and the corresponding coalescence parameter. At the end of the process we get $D_h \approx -30\, \textrm{dB}$ (dB stands for $10\,\textrm{log}_{10}(D_h)$). 

\section{Appendix B: Calculating the threshold gain}
\begin{figure}[H]
    \centering
    \includegraphics[scale=0.6]{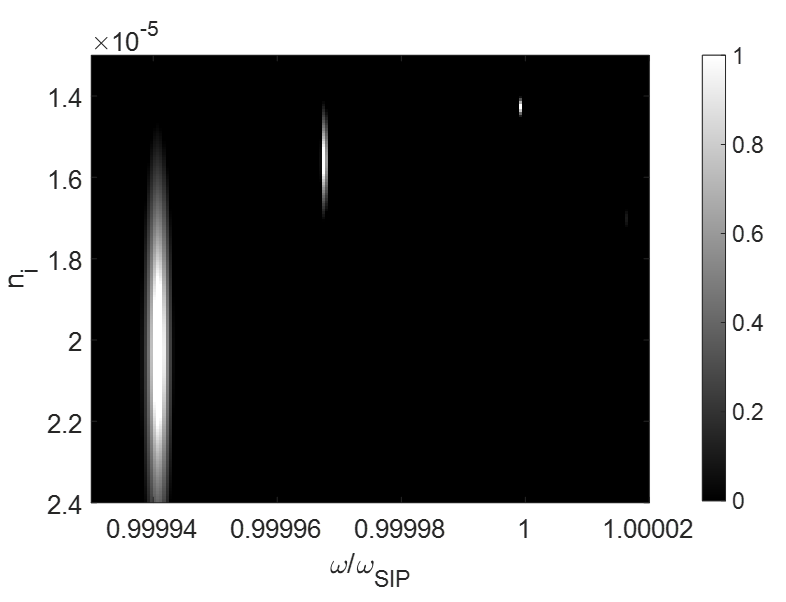}
    \caption{Log scale of the transmission power for a finite-length structure with $56$ unit cells, and single Input port and transmission port. The transmission is plotted as a function of the angular frequency and the imaginary part of the effective refractive index of a uniform (single mode) waveguide. Three lasing conditions are visible as the narrow frequency sharp peaks.}
    \label{fig:2D_transmission_gain}
\end{figure}

In this appendix, we present the process for calculating the threshold gain level and the resonance frequency for a finite-length structure with single input and transmission ports, as depicted in Fig.~\ref{fig:closed_model}. A uniformly distributed gain is added to the structure, by introducing an imaginary part ($n_i$) to the refractive index of the waveguides and the ring resonators. 
First, we find the amplitude transmission of the structure as a function of the frequency and gain. We then plot the absolute value of the amplitude transmission $\textrm{Log}_{10} |T^2|$ as a function of these parameters, as depicted in Fig.~\ref{fig:2D_transmission_gain} for a structure comprising $56$ unit cells (in logarithmic scale). 
This plot allows for the identification of the resonance (i.e., lasing) frequencies {\em and} and their corresponding lasing thresholds. A threshold condition corresponds to a pole in the transmission function, which is manifested as a peak in the transmission function. 
Next, we identify the resonance frequency closest to the SIP frequency (the SIP frequency was obtained by using the optimization process described in Appendix A). While the resonance frequency does not coincide with the SIP frequency, it approaches it for a very large number of unit cells, i.e., $\omega_{res}\xrightarrow{N\rightarrow\infty } \omega_{SIP}$ \cite{Giant_resonances_photonic_crystals2010}. The gain level at which the transmission is maximal is the threshold gain of the structure at the relative resonance (i.e., lasing) frequency.
The lasing threshold is identified as the gain at which the transmission is maximal at the resonance frequency of interest.

\bibliography{Bib_SIP_3w1r}

\end{document}